\def\Statusstring{Submitted on April 27, 2006}

\documentclass[11pt,letter]{article}

\usepackage{amsmath,amscd,fancyhdr,graphics,pifont}
\usepackage{floatflt,subfigure,epsfig,moreverb,multicol,lscape}
\usepackage{supertabular,tabularx,multirow,bar,amssymb}
\usepackage{theorem,psfrag,afterpage,curves,epic, bbm}
\usepackage[english]{babel}

\newcommand{\ignore}[1]{}

\newcommand{\matr}[1]{\mathbf{#1}}
\newcommand{\vect}[1]{\mathbf{#1}}

\newcommand{\GF}[1]{\mathbb{F}_{#1}}

\newcommand{\defeq}{\triangleq}

\newcommand{\vu}{\vect{u}}
\newcommand{\vx}{\vect{x}}

\newcommand{\vy}{\vect{y}}

\newcommand{\tr}{\mathsf{T}}

\newcommand{\HD}[2]{\mathcal{D}_{#1}^{(#2)}}
\newcommand{\iL}{\mathsf{L}}
\newcommand{\iX}{\mathsf{X}}

\newcommand{\us}{\upsilon} 
\newcommand{\usSet}{\Upsilon} 

\makeatletter
  \def\revddots{\mathinner{\mkern1mu\raise\p@
  \vbox{\kern7\p@\hbox{.}}\mkern2mu
  \raise4\p@\hbox{.}\mkern2mu\raise7\p@\hbox{.}\mkern1mu}}
  \makeatother

\newtheorem{Lemma}{Lemma}

\newtheorem{Proposition}[Lemma]{Proposition}
\newtheorem{Corollary}[Lemma]{Corollary}

\theoremstyle{plain}

\newtheorem{PreDefinition}[Lemma]{{\textbf{Definition}}}
  \newenvironment{Definition}%
    {\begin{PreDefinition}}{\hfill$\square$\end{PreDefinition}}

\theoremstyle{plain}

\newtheorem{PreAlgorithm}[Lemma]{{\textbf{Algorithm}}}
  \newenvironment{Algorithm}%
    {\begin{PreAlgorithm}\upshape}{\hfill$\square$\end{PreAlgorithm}}

\newtheorem{PreRemark}[Lemma]{{\textbf{Remark}}}
  \newenvironment{Remark}%
    {\begin{PreRemark}\upshape}{\hfill$\square$\end{PreRemark}}

\newtheorem{PreExample}[Lemma]{{\textbf{Example}}}
  \newenvironment{Example}%
    {\begin{PreExample}\upshape}{\hfill$\square$\end{PreExample}}

\newenvironment{Proof}%
  {\noindent \emph{Proof:}}{\hfill$\square$}

\begin{document}

\title{On Universally Decodable Matrices for Space-Time Coding%
  \footnote{The first author was supported by NSF Grants ATM-0296033 and DOE
            SciDAC and by ONR Grant N00014-00-1-0966. The second author was
            supported by NSF Grant CCF-0514801. Some results discussed here
            also appeared in~\cite{Ganesan:Boston:05:1,
            Ganesan:Vontobel:06:1:subm, Vontobel:Ganesan:06:1}.}}

\author{Pascal O.~Vontobel%
  \thanks{Department of EECS, Massachusetts Institute of Technology,
          77 Massachusetts Avenue, Cambridge, MA 02139 USA.
          Email: \texttt{pascal.vontobel@ieee.org}. P.O.V.~is the
          corresponding author.}
  \ and Ashwin Ganesan%
  \thanks{ECE Department, University of
          Wisconsin-Madison, 1415 Engineering Drive Madison, WI 53706, USA.
          Email: \texttt{ganesan@cae.wisc.edu}.}}

\date{}

\maketitle

\vspace{-6cm}
\begin{flushright}
  \texttt{\Statusstring}\\[1cm]
\end{flushright}
\vspace{+4cm}

\begin{abstract}
  The notion of universally decodable matrices (UDMs) was recently introduced
  by Tavildar and Viswanath while studying slow fading channels. It turns out
  that the problem of constructing UDMs is tightly connected to the problem of
  constructing maximum distance separable (MDS) codes. In this paper, we first
  study the properties of UDMs in general and then we discuss an explicit
  construction of a class of UDMs, a construction which can be seen as an
  extension of Reed-Solomon codes. In fact, we show that this extension is, in
  a sense to be made more precise later on, unique. Moreover, the structure of
  this class of UDMs allows us to answer some open conjectures by Tavildar,
  Viswanath, and Doshi in the positive, and it also allows us to formulate an
  efficient decoding algorithm for this class of UDMs. It turns out that our
  construction yields a coding scheme that is essentially equivalent to a
  class of codes that was proposed by Rosenbloom and Tsfasman. Moreover, we
  point out connections to so-called repeated-root cyclic codes.
\end{abstract}

\mbox{} \\

\noindent\textbf{2000 Mathematics Subject Classification:} 
                 05B20, 15A03, 94B05, 94B35 \\

\noindent\textbf{Index terms} --- Universally decodable matrices, coding
  for slow fading channels, rank condition, Rosenbloom-Tsfasman metric,
  Reed-Solomon codes, repeated-root cyclic codes, Hasse derivatives, Newton
  interpolation, Pascal's triangle.

\newpage

\section{Introduction}
\label{sec:introduction:1}

\begin{figure}[h]
  \begin{center}
    \epsfig{file=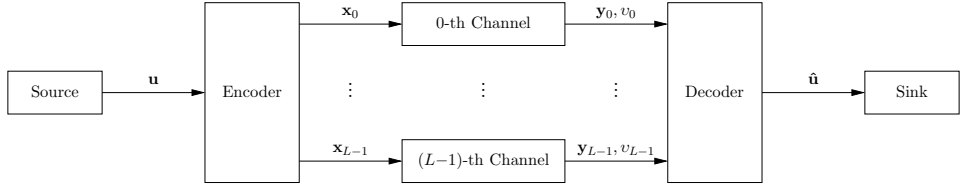, width=12cm}
  \end{center}
  \caption{Communication system with $L$ parallel channels.}
  \label{fig:communication:system:1}
\end{figure}

Let $L$, $N$, and $K$ be positive integers, let $q$ be a prime power, let $[M]
\defeq \{ 0, \ldots, M-1 \}$ for any positive integer $M$, and let $[M] \defeq
\{ \ \}$ for any non-positive integer $M$. While studying slow fading
channels (c.f.~e.g.~\cite{Tse:Viswanath:05:1}), Tavildar and
Viswanath~\cite{Tavildar:Viswanath:05:2} introduced the communication system
shown in Fig.~\ref{fig:communication:system:1} which works as follows. An
information (column) vector $\vect{u} \in \GF{q}^K$ is encoded into codeword
vectors $\vx_{\ell} \defeq \matr{A}_{\ell} \cdot \vect{u} \in
\GF{q}^{N}$, $\ell \in [L]$, where $\matr{A}_0, \ldots, \matr{A}_{L-1}$ are $L$
matrices over $\GF{q}$ and of size $N \times K$. (Actually, Tavildar and
Viswanath~\cite{Tavildar:Viswanath:05:2} considered only the special case $K =
N$.) Upon sending $\vx_{\ell}$ over the $\ell$-th channel we receive
$\vy_{\ell} \in (\GF{q} \cup \{ ? \})^N$, where the question mark denotes an
erasure. The channels are such that the received vectors $\vect{y}_0,
\ldots, \vect{y}_{L-1}$ can be characterized as follows: there are integers
$\us_0, \ldots, \us_{L-1}$, $0 \leq \us_{\ell}
\leq N$, $\ell \in [L]$ (that can vary from transmission to transmission) such
that the first $\us_{\ell}$ entries of $\vect{y}_{\ell}$ are non-erased and
agree with the corresponding entries of $\vect{x}_{\ell}$ and such that the
last $N - \us_{\ell}$ entries of $\vect{y}_{\ell}$ are erased.

Based on the non-erased entries we would like to reconstruct
$\vect{u}$. The obvious decoding approach works as follows:
construct a $(\sum_{\ell \in [L]} \us_{\ell}) \times K$-matrix
$\matr{A}$ that stacks the $\us_0$ first rows of $\matr{A}_0$,
$\ldots$, the $\us_{L-1}$ first rows of $\matr{A}_{L-1}$; then
construct a length-$(\sum_{\ell \in [L]} \us_{\ell})$ vector
$\vect{y}$ that concatenates the $\us_0$ first entries of
$\matr{y}_0$, $\ldots$, the $\us_{L-1}$ first entries of
$\vect{y}_{L-1}$; finally, the vector $\vect{\hat u}$ is given as
the solution of the linear equation system $\matr{A} \cdot
\vect{\hat u} = \vect{y}$. Since $\vect{u}$ is arbitrary in
$\GF{q}^K$, a necessary condition for successful decoding is that
$\sum_{\ell \in [L]} \us_{\ell} \geq K$. Because we would like to be
able to decode correctly for all $L$-tuples $(\us_0, \ldots,
\us_{L-1})$ that satisfy this necessary condition, we must guarantee
that the matrix $\matr{A}$ has full rank for all possible $L$-tuples
$(\us_0, \ldots, \us_{L-1})$ with $\sum_{\ell \in [L]} \us_{\ell}
\geq K$. Matrices that fulfill this condition are called universally
decodable matrices (UDMs) and will be formally defined in
Sec.~\ref{sec:udm:1}.

There is a tight connection between UDMs and maximum-distance separable (MDS)
codes~\cite[Ch.~11]{MacWilliams:Sloane:77}. Indeed, assume that $L
\geq K$ and consider the $L \times K$-matrix $\matr{G}$ that consists of the
zeroth row of $\matr{A}_0$, $\ldots$, the zeroth row of $\matr{A}_{L-1}$.
Looking at all the cases where $(\us_0, \ldots, \us_{L-1})$ is such that
$\sum_{\ell \in [L]} \us_{\ell} = K$ and such that $0 \leq \us_{\ell} \leq 1$
for all $\ell \in [L]$, we see that all $K \times K$ sub-matrices of
$\matr{G}$ must have full rank. Now, Th.~1 in
~\cite[Ch.~11]{MacWilliams:Sloane:77} implies that $\matr{G}$ must be the
generator matrix of a $q$-ary MDS code of length $L$ and dimension $K$.

Given the definition of UDMs, there are several immediate questions.  For what
values of $L$, $N$, $K$, and $q$ do such matrices exist?  What are the
properties of these matrices? How can one construct such matrices?
In~\cite{Tavildar:Viswanath:05:2} a construction is given for $L = 3$, any
$N$, $K = N$, and $q = 2$.  Doshi~\cite{Doshi:05:1} gave a construction for $L
= 4$, $N = K = 3$, and $q = 3$ and conjectured a construction for $L = 4$, $N$
any power of $3$, $K = N$, and $q = 3$. Ganesan and
Boston~\cite{Ganesan:Boston:05:1} showed that for any $N \geq 2$, $K = N$, the
value $L$ is upper bounded by $L \leq q+1$, and conjectured that this
condition is also sufficient. The correctness of this conjecture was
subsequently proved in~\cite{Vontobel:Ganesan:05:1,
Ganesan:Vontobel:06:1:subm}. In this paper we generalize this bound to the
case $K \leq 2N$ and we will give an explicit construction that works for any
positive integers $L$, $N$, $K$ and any prime power $q$ as long as $L \leq
q+1$, in other words, this construction achieves for any $K \leq 2N$, $N \geq
2$, and any prime power $q$ the above-mentioned upper bound on $L$.  As a side
result, our construction shows that the above-mentioned conjecture by Doshi is
indeed true. We will also show that for $K = N$ this construction is (in a
sense to be made more precise) the uniquely possible way to extend a
Reed-Solomon code (which is an MDS code) to UDMs. Finally, we will present an
efficient decoding algorithm for the UDMs given by the above-mentioned
construction, i.e.~we will present an algorithm that efficiently solves
$\matr{A} \cdot
\vect{\hat u} = \vect{y}$.

We will point out several connections to other codes. As already mentioned,
there is a tight connection between UDMs and MDS codes, but we will also point
out an interesting relationship to so-called repeated-root cyclic
codes. Moreover, it turns out that the above-mentioned construction of UDMs is
essentially equivalent to so-called Reed-Solomon $m$-codes, a class of codes
described by Rosenbloom and Tsfasman~\cite[Sec.~3]{Rosenbloom:Tsfasman:97:1}.
These authors were interested in coding under a non-Hamming metric, namely a
metric they called the $m$-metric and that is now also known as the
Rosenbloom-Tsfasman metric. For this metric, Rosenbloom and Tsfasman show that
the Reed-Solomon $m$-codes achieve the Singleton bound.

The paper is structured as follows. In Sec.~\ref{sec:udm:1} we properly define
UDMs and in Sec.~\ref{sec:modifying:udms:1} we show how UDMs can be modified
to obtain new UDMs.  Sec.~\ref{sec:explicit:construction:udms:1} is the main
section where an explicit construction of UDMs is presented and in
Sec.~\ref{sec:decoding:1} we discuss an efficient decoding algorithm for these
UDMs. In Sec.~\ref{sec:conclusions:1} we offer some conclusions. Finally,
Sec.~\ref{sec:proofs:1} contains the longer proofs and
Sec.~\ref{sec:hasse:derivatives:1} collects some results on Hasse derivatives
which are the main tool for the proof of our UDMs construction.

\section{Universally Decodable Matrices}
\label{sec:udm:1}

The notion of universally decodable matrices (UDMs) was introduced by Tavildar
and Viswanath~\cite{Tavildar:Viswanath:05:2}. Before giving the definition of
UDMs, let us agree on some notation. For any positive integer $K$, we let
$\matr{I}_K$ be the $K \times K$ identity matrix and we let $\matr{J}_K$ be
the $K \times K$ matrix where all entries are zero except for the
anti-diagonal entries that are equal to one; i.e., $\matr{J}_K$ contains the
rows of $\matr{I}_K$ in reverse order. For any positive integers $N$ and $K$
with $N \leq K$ we let $\matr{I}_{N,K}$ and $\matr{J}_{N,K}$ be the first $N$
rows of $\matr{I}_K$ and $\matr{J}_K$, respectively. Row and column indices of
matrices will always be counted from zero on and the entry in the $i$-th row
and $j$-th column of a matrix $\matr{A}$ will be denoted by
$[\matr{A}]_{i,j}$. Similarly, indices of vectors will be counted from zero on
and the $i$-th entry of a vector $\vect{a}$ will be denoted by
$[\vect{a}]_i$. For any positive integer $L$, $N$, and $K$ we define the sets
\begin{align*}
  \usSet^{= K}_{L,N}
    &\defeq
       \left\{
           (\us_0, \ldots, \us_{L-1})
         \
         \left|
         \
           0 \leq \us_{\ell} \leq N,
           \ell \in [L], \
           \sum_{\ell \in [L]} \us_{\ell} = K
         \right.
       \right\}, \\
  \usSet^{\geq K}_{L,N}
    &\defeq
       \left\{
           (\us_0, \ldots, \us_{L-1})
         \
         \left|
         \
           0 \leq \us_{\ell} \leq N,
           \ell \in [L], \
           \sum_{\ell \in [L]} \us_{\ell} \geq K
         \right.
       \right\}.
\end{align*}

\begin{Definition}
  \label{def:udms:1}

  Let $N$, $K$, and $L$ be some positive integers and let $q$ be a prime
  power. The $L$ matrices $\matr{A}_0, \ldots, \matr{A}_{L-1}$ over $\GF{q}$
  and of size $N \times K$ are $(L,N,K,q)$-UDMs, or simply UDMs, if for every
  $(\us_0, \ldots, \us_{L-1}) \in \usSet^{\geq K}_{L,N}$ they fulfill the UDMs
  condition which says that the $( \sum_{\ell \in [L]} \us_{\ell} ) \times K$
  matrix composed of the first $\us_0$ rows of $\matr{A}_0$, the first $\us_1$
  rows of $\matr{A}_1$, $\ldots$, the first $\us_{L-1}$ rows of
  $\matr{A}_{L-1}$, has full rank.
\end{Definition}

In the following we will only consider $(L,N,K,q)$-UDMs for which $N \leq K
\leq LN$ holds. The reason for the first inequality is that for the purpose of
unique decodability it does not help to send more than $K$ symbols over the
$\ell$-th channel, $\ell \in [L]$. (This condition might be weakened though
for channel models that introduce not only erasures but also errors.)  The
reason for the second inequality is that if $K > LN$ then we will never
receive enough symbols to decode uniquely. Note that for $K = N$, i.e.~the
case studied by Tavildar and Viswanath~\cite{Tavildar:Viswanath:05:2}, both
conditions in $N \leq K \leq LN$ are fulfilled for any positive $L$.

We list some immediate consequences of the above definition.
\begin{itemize}

  \item To assess that some matrices $\matr{A}_0, \ldots, \matr{A}_{L-1}$ are
    UDMs, it is sufficient to check the UDMs condition only for every $(\us_0,
    \ldots, \us_{L-1}) \in \usSet^{= K}_{L,N}$. In the case $K = N$ there are
    ${ N+L-1 \choose L-1 }$ such $L$-tuples.

  \item If the matrices $\matr{A}_0, \ldots, \matr{A}_{L-1}$ are UDMs then
    all these matrices have full rank.

  \item If the matrices $\matr{A}_0, \ldots, \matr{A}_{L-1}$ are
    $(L,N,K,q)$-UDMs then they are $(L,N,K,q')$-UDMs for any $q'$ that is a
    power of $q$.

  \item Let $\sigma$ be any permutation of $[L]$. If the matrices $\matr{A}_0,
    \ldots, \matr{A}_{L-1}$ are $(L,N,K,q)$-UDMs then the matrices
    $\matr{A}_{\sigma(0)}, \ldots, \matr{A}_{\sigma(L-1)}$ are also
    $(L,N,K,q)$-UDMs.

  \item If the matrices $\matr{A}_0, \ldots, \matr{A}_{L-1}$ are
    $(L,N,K,q)$-UDMs then the matrices $\matr{A}_0, \ldots, \matr{A}_{L'-1}$
    are $(L',N,K,q)$-UDMs for any positive $L'$ with $L' \leq L$. (Note that
    the condition $K \leq L'N$ may be violated.)

  \item If the matrices $\matr{A}_0, \ldots, \matr{A}_{L-1}$ are
    $(L,N,K,q)$-UDMs and $\matr{B}$ is an invertible $K \times K$-matrix over
    $\GF{q}$ then the matrices $\matr{A}_0 \cdot \matr{B}, \ldots,
    \matr{A}_{L-1} \cdot \matr{B}$ are $(L,N,K,q)$-UDMs. Without loss of
    generality, we can therefore assume that $\matr{A}_0 = \matr{I}_{N,K}$.

  \item For $K = 1$ (note that we must have $N = 1$ because we assume that $N
    \leq K$) we see that for any positive integer $L$ and any prime power $q$,
    the $L$ matrices $(1), \ldots, (1)$ are $(L,N{=}1,K{=}1,q)$-UDMs. Because
    of the triviality of the case $K = 1$, the rest of the paper focuses on
    the case $K \geq 2$.

\end{itemize}

\begin{Example}
  \label{ex:udm:1}

  Let $N$ be any positive integer, let $q$ be any prime power, let $L \defeq
  2$, let $\matr{A}_0 \defeq \matr{I}_N$ and let $\matr{A}_1 \defeq
  \matr{J}_N$. It can easily be checked that $\matr{A}_0, \matr{A}_1$ are
  $(L{=}2,N,K{=}N,q)$-UDMs.

  Let us verify this statement for $N \defeq 5$: we must check that for any
  non-negative integers $\us_1$ and $\us_2$ such that $\us_1 + \us_2 = 5$ the
  UDMs condition is fulfilled, which in the case $(\us_1, \us_2) = (3, 2)$
  means that we must show that the matrix
  \begin{align*}
    \begin{pmatrix}
      1 & 0 & 0 & 0 & 0 \\
      0 & 1 & 0 & 0 & 0 \\
      0 & 0 & 1 & 0 & 0 \\
      0 & 0 & 0 & 0 & 1 \\
      0 & 0 & 0 & 1 & 0
    \end{pmatrix}
  \end{align*}
  has rank $5$. This is easily done. For the other $(\us_1, \us_2)$-tuples in
  question the verification is done equally easily.
\end{Example}

\begin{Example}
  \label{ex:udm:2}

  In order to give the reader a feeling how UDMs might look like for $L > 2$,
  we give here a simple example for $L = 4$, $N = K = 3$, and $q = 3$, namely
  \begin{align*}
    \matr{A}_0
      &= \begin{pmatrix}
           1 & 0 & 0 \\
           0 & 1 & 0 \\
           0 & 0 & 1
         \end{pmatrix},
         \quad
    \matr{A}_1
       =
         \begin{pmatrix}
           0 & 0 & 1 \\
           0 & 1 & 0 \\
           1 & 0 & 0
         \end{pmatrix},
         \quad
    \matr{A}_2
       =
         \begin{pmatrix}
           1 & 1 & 1 \\
           0 & 1 & 2 \\
           0 & 0 & 1
         \end{pmatrix},
         \quad
    \matr{A}_3
       =
         \begin{pmatrix}
           1 & 2 & 1 \\
           0 & 1 & 1 \\
           0 & 0 & 1
         \end{pmatrix}.
  \end{align*}
  One can verify that for all $(\us_0, \us_1, \us_2, \us_3) \in \usSet^{=
  3}_{4,3}$ (there are $20$ such four-tuples) the UDMs condition is fulfilled
  and hence the above matrices are indeed UDMs. For example, for
  $(\us_0,\us_1,\us_2,\us_3) = (0,0,3,0)$, $(\us_0,\us_1,\us_2,\us_3) =
  (0,0,1,2)$, and $(\us_0,\us_1,\us_2,\us_3) = (1,1,0,1)$ the UDMs condition
  means that we have to check if the matrices
  \begin{align*}
    \begin{pmatrix}
      1 & 1 & 1 \\
      0 & 1 & 2 \\
      0 & 0 & 1
    \end{pmatrix},
    \quad
    \begin{pmatrix}
      1 & 1 & 1 \\
      1 & 2 & 1 \\
      0 & 1 & 1
    \end{pmatrix},
    \quad
    \begin{pmatrix}
      1 & 0 & 0 \\
      0 & 0 & 1 \\
      1 & 2 & 1
    \end{pmatrix}
  \end{align*}
  have rank $3$, respectively, which is indeed the case. Before concluding
  this example, let us remark that the above UDMs are the same UDMs that
  appeared in~\cite{Doshi:05:1}
  and~\cite[Sec.~4.5.4]{Tavildar:Viswanath:05:2}.
\end{Example}

\section{Modifying UDMs}
\label{sec:modifying:udms:1}

This section discusses ways to modify UDMs such that new UDMs result. Besides
the intrinsic interest in such results, the insights that we gain can be used
towards deriving some necessary conditions for the existence of UDMs (see
Lemmas~\ref{lemma:maximal:L:1} and
\ref{lemma:maximal:L:2}).

\begin{Lemma}
  \label{lemma:replace:and:add:lines:1}

  Let $\matr{A}_0, \ldots, \matr{A}_{L-1}$ be $(L,N,K,q)$-UDMs. For any $\ell
  \in [L]$ and $n \in [N]$ we can replace the $n$-th row of $\matr{A}_{\ell}$
  by any non-zero multiple of itself without violating any UDMs
  condition. Moreover, for any $\ell \in [L]$ and $n, n' \in [N]$, $n > n'$,
  we can add any multiples of the $n'$-th row of $\matr{A}_{\ell}$ to the
  $n$-th row of $\matr{A}_{\ell}$ without violating any UDMs condition. More
  generally, the matrix $\matr{A}_{\ell}$ can be replaced by $\matr{C}_{\ell}
  \cdot \matr{A}_{\ell}$ without violating any UDMs condition, where
  $\matr{C}_{\ell}$ is an arbitrary lower triangular $N \times N$-matrix over
  $\GF{q}$ with non-zero diagonal entries.
\end{Lemma}

\begin{Proof}
  Follows from well-known properties of determinants.
\end{Proof}

\begin{Lemma}
  \label{lemma:tensoring:1}

  Let $\matr{A}_0, \ldots, \matr{A}_{L-1}$ be $(L,N,K,q)$-UDMs for which we
  know that the tensor powers $\matr{A}_0^{\otimes m}, \ldots,
  \matr{A}_{L-1}^{\otimes m}$ are $(L,N^m,K^m,q)$-UDMs for some positive
  integer $m$. For all $\ell \in [L]$, let $\matr{A}'_{\ell} \defeq
  \matr{A}_{\ell} \cdot \matr{B}$, where $\matr{B}$ is an arbitrary invertible
  $K \times K$ matrix over $\GF{q}$. Then $\matr{A}'_0, \ldots,
  \matr{A}'_{L-1}$ are $(L,N,K,q)$-UDMs and $(\matr{A}'_0)^{\otimes m},
  \ldots, (\matr{A}'_{L-1})^{\otimes m}$ are $(L,N^m,K^m,q)$-UDMs. On the
  other hand, if for all $\ell \in [L]$ we define $\matr{A}'_{\ell} \defeq
  \matr{C}_{\ell} \cdot \matr{A}_{\ell}$, where $\matr{C}_{\ell}$, $\ell \in
  [L]$, are lower-triangular matrices with non-zero diagonal entries, then
  $\matr{A}'_0, \ldots, \matr{A}'_{L-1}$ are $(L,N,K,q)$-UDMs and
  $(\matr{A}'_0)^{\otimes m}, \ldots, (\matr{A}'_{L-1})^{\otimes m}$ are
  $(L,N^m,K^m,q)$-UDMs.
\end{Lemma}

\begin{Proof}
  This follows from the sixth comment after Def.~\ref{def:udms:1}, from
  Lemma~\ref{lemma:replace:and:add:lines:1}, and by using a well-known
  property of tensor products, namely that $(\matr{M}_{1} \cdot
  \matr{M}_2)^{\otimes m} = \matr{M}_1^{\otimes m} \cdot \matr{M}_2^{\otimes
  m}$ for any compatible matrices $\matr{M}_1$ and $\matr{M}_2$. Note that
  $\matr{C}_{\ell}^{\otimes m}$ is a lower-triangular matrix with non-zero
  diagonal entries for all $\ell \in [L]$ and positive integers $m$.
\end{Proof}

\begin{Lemma}
  \label{lemma:reversed:order:2}

  Let $\matr{A}_0, \ldots, \matr{A}_{L-1}$ be $(L,N,K,q)$-UDMs. Then there
  exist matrices $\matr{A}'_0, \ldots, \matr{A}'_{L-1}$ that are
  $(L,N,K,q)$-UDMs and where for all $\ell' \in \lfloor L/2 \rfloor$ the the
  $n$-th row of $\matr{A}'_{2\ell'}$ equals the $(K{-}1{-}n)$-th row of
  $\matr{A}'_{2\ell'+1}$ for all $K-N \leq n \leq N-1$. In the case $K = N$
  this means that for all $\ell' \in \lfloor L/2 \rfloor$ the matrix
  $\matr{A}'_{2\ell'+1}$ is the same as $\matr{A}'_{2\ell'}$ except that the
  rows are in reversed order, i.e.~$\matr{A}_{2\ell'+1} = \matr{J}_N \cdot
  \matr{A}'_{2\ell'}$.
\end{Lemma}

\begin{Proof}
  See Sec.~\ref{sec:proof:lemma:reversed:order:2}.
\end{Proof}

\begin{Remark}
  \label{remark:identity:structure:1}

  From Lemma~\ref{lemma:reversed:order:2} we see that when considering
  $(L,N,K,q)$-UDMs $\matr{A}_0, \ldots, \matr{A}_{L-1}$ we can without loss of
  generality assume that $\matr{A}_0 = \matr{I}_{N,K}$ and that $\matr{A}_1 =
  \matr{J}_{N,K}$.
\end{Remark}

\begin{Proof}
  See Sec.~\ref{sec:proof:remark:identity:structure:1}.
\end{Proof}

\begin{Lemma}
  \label{lemma:reduction:1}

  Let the matrices $\matr{A}_0, \ldots, \matr{A}_{L-1}$ be $(L,N,K,q)$-UDMs
  with $\matr{A}_0 = \matr{I}_{N,K}$ and $\matr{A}_1 = \matr{J}_{N,K}$. The
  matrices $\matr{A}'_0, \ldots, \matr{A}'_{L-1}$ are $(L,N{-}1,K{-}1,q)$-UDMs
  if $\matr{A}'_{\ell}$ is obtained as follows from $\matr{A}_{\ell}$: if
  $\ell = 1$ then delete the first column and last row of $\matr{A}_{\ell}$,
  otherwise delete the last column and last row of $\matr{A}_{\ell}$. (Note
  that possibly the new UDMs do not fulfill $(N-1)L \geq K-1$.)
\end{Lemma}

\begin{Proof}
  See Sec.~\ref{sec:proof:lemma:reduction:1}.
\end{Proof}

\mbox{}

The following two lemmas show how the above lemmas can be used to obtain upper
bounds on $L$.

\begin{Lemma}
  \label{lemma:maximal:L:1}

  Let $q$ be a prime power. If $K \geq 2$ then $(L,N,K{\leq}2N,q)$-UDMs can
  only exist for $L \leq q + 1$. (Note that this upper bound on $L$ is
  independent of $N$ and $K$ as long as $2 \leq K \leq 2N$.)
\end{Lemma}

\begin{Proof}
  See Sec.~\ref{sec:proof:lemma:maximal:L:1}.
\end{Proof}

\begin{Lemma}
  \label{lemma:maximal:L:2}

  Let $q$ be a prime power. Then $(L,N,K{=}2N{+}1,q)$-UDMs can only exist for
  $L \leq q + 2$. (Note that this upper bound on $L$ is independent of $N$.)
\end{Lemma}

\begin{Proof}
  See Sec.~\ref{sec:proof:lemma:maximal:L:2}.
\end{Proof}

\mbox{}

Note that Lemmas~\ref{lemma:maximal:L:1} and~\ref{lemma:maximal:L:2} are
generalizations of a result in~\cite{Ganesan:Boston:05:1} that dealt with the
case $K = N$. Prop.~\ref{prop:udms:construction:1} will show that the upper
bound on $L$ in Lemma~\ref{lemma:maximal:L:1} is the best possible because for
any $L \leq q+1$ we can explicitly construct
$(L,N,K{\leq}2N,q)$-UDMs. Moreover, as Rem.~\ref{remark:special:mds:code:1}
shows, the upper bound on $L$ in Lemma~\ref{lemma:maximal:L:2} is the best
possible if no further restrictions on $N$ and $q$ are imposed.

\begin{Remark}
  \label{remark:special:mds:code:1}

  Let $q \defeq 2^s$ for some integer $s$ and let $\alpha$ be a primitive
  element in $\GF{q}$, i.e.~$\alpha$ is a $(q-1)$-th primitive root of
  unity. From~\cite[Th.~10 in Ch.~11]{MacWilliams:Sloane:77} it follows that
  the matrices
  \begin{align*}
    \matr{A}_0
      &\defeq \begin{pmatrix}
                1 & 0 & 0
              \end{pmatrix}, \\
    \matr{A}_1
      &\defeq \begin{pmatrix}
                0 & 0 & 1
              \end{pmatrix}, \\
    \matr{A}_{\ell+2}
      &\defeq \begin{pmatrix}
                1 & \alpha^{\ell} & \alpha^{2\ell}
              \end{pmatrix}
              \quad
              \text{for $0 \leq \ell \leq q-2$}, \\
    \matr{A}_{q+1}
      &\defeq \begin{pmatrix}
                0 & 1 & 0
              \end{pmatrix}
  \end{align*}
  are $(q{+}2,1,3,q)$-UDMs. 
\end{Remark}

For $K > 2N + 1$ it is more complicated to find an upper bound on $L$ in terms
of $q$. In particular, for $N = 1$ the question of finding upper bounds on $L$
is equivalent to the question of the existence of MDS
codes~\cite[Ch.~11]{MacWilliams:Sloane:77}: it is conjectured that for $2 \leq
K \leq L-2$ we must have $L \leq q+1$. (The only known exception to this
conjecture are MDS codes of length $L = q+2$ and of dimension $K = 3$ or $K =
L-3$ for $q = 2^s$ where $s$ is some positive integer. For the $K = 3$ case,
see Rem.~\ref{remark:special:mds:code:1}, for the $K = L-3$ case,
see~\cite[Th.~10 in Ch.~11]{MacWilliams:Sloane:77}.)

\section{An Explicit Construction of UDMs}
\label{sec:explicit:construction:udms:1}

In this section we would like to present an explicit construction of
$(L,N,K,q)$-UDMs, cf.~Prop.~\ref{prop:udms:construction:1} and
Cor.~\ref{cor:prop:udms:construction:1}. This construction is very much
motivated by the connection of UDMs to MDS codes mentioned in
Sec.~\ref{sec:introduction:1} and the fact that Reed-Solomon codes are MDS
codes. In fact, we will see in Prop.~\ref{prop:udms:construction:uniqueness:1}
that there is (in a sense to be made more precise) only one possible way to
construct UDMs based on Reed-Solomon codes.

Before we proceed, we need some definitions. First, whenever necessary we use
the natural mapping of the integers into the prime subfield\footnote{When $q =
p^s$ for some prime $p$ and some positive integer $s$ then $\GF{p}$ is a
subfield of $\GF{q}$ and is called the prime subfield of $\GF{q}$. $\GF{p}$
can be identified with the integers $\{ 0, 1, \ldots, p-1 \}$, where addition
and multiplication are modulo $p$.} of $\GF{q}$. Secondly, we define the
binomial coefficient ${a
\choose b}$ in the usual way. Note that ${a \choose b} = 0$ for
all $a < b$.

\begin{Definition}
  \label{def:taylor:polynomial:expansion:1}

  Let $a(\iX) \defeq \sum_{k=0}^{d} a_k \iX^k \in \GF{q}[\iX]$ be a polynomial
  and let $\beta \in \GF{q}$. The Taylor polynomial expansion of $a(\iX)$
  around $\iX = \beta$ is defined to be $a(\iX) = \sum_{n=0}^{d} a_{\beta,n}
  (\iX - \beta)^n \in \GF{q}[\iX]$ for suitably chosen $a_{\beta,n} \in
  \GF{q}$, $0 \leq n \leq d$, such that equality holds.
\end{Definition}

It can be verified that the Taylor polynomial coefficients $a_{\beta,n}$ can
be expressed using Hasse derivatives\footnote{See
Sec.~\ref{sec:hasse:derivatives:1} for the definition and some properties of
Hasse derivatives.} of $a(\iX)$, i.e.~$a_{\beta,n} = a^{(n)}(\beta) =
\sum_{k=0}^{d} a_k {k \choose n} \beta^{k-n}$. On the other hand, the
coefficients of $a(\iX)$ can be expressed as $a_k = \sum_{n=0}^{d} a_{\beta,n}
{n \choose k} (-\beta)^{n-k}$.

\begin{Lemma}
  \label{lemma:zero:multiplicity:1}

  Let $a(\iX) \defeq \sum_{k=0}^{d} a_k \iX^k \in \GF{q}[\iX]$ be a non-zero
  polynomial, let $\beta \in \GF{q}$, and let $a(\iX) = \sum_{n=0}^{d}
  a_{\beta,n} (\iX - \beta)^n \in \GF{q}[\iX]$ be the Taylor polynomial
  expansion of $a(\iX)$ around $\iX = \beta$. The polynomial $a(\iX)$ has a
  zero at $\iX = \beta$ of multiplicity $m$ if and only if $a_{\beta,n} = 0$
  for $0 \leq n < m$ and $a_{\beta,m} \neq 0$.
\end{Lemma}

\begin{Proof}
  Obvious.
\end{Proof}

\mbox{}

In the following, evaluating the $n$-th Hasse derivative $u^{(n)}(\iL)$ of a
polynomial $u(\iL)$ at $\iL = \infty$ shall result in the value $u_{K-1-n}$,
i.e.~we set $u^{(n)}(\infty) \defeq u_{K-1-n}$.

\begin{Proposition}
  \label{prop:udms:construction:1}

  Let $N$ and $K$ be some positive integers, let $q$ be some prime power, and
  let $\alpha$ be a primitive element in $\GF{q}$. If $L \leq q+1$ then the
  following $L$ matrices over $\GF{q}$ of size $N \times K$ are
  $(L,N,K,q)$-UDMs:
  \begin{align*}
    &
    \matr{A}_0
       \defeq \matr{I}_{N,K},
    \quad
    \matr{A}_1
       \defeq \matr{J}_{N,K},
    \quad
    \matr{A}_{2},
    \quad
    \ldots,
    \quad
    \matr{A}_{L-1}, \\
    &
    \text{ where } \
    [\matr{A}_{\ell+2}]_{n,k}
       \defeq {k \choose n} \alpha^{\ell (k-n)}, \
              (\ell,n,k) \in [L-2] \times [N] \times [K].
  \end{align*}
  Note that ${k \choose n}$ is to be understood as follows: compute ${k
  \choose n}$ over the integers and apply only then the natural mapping to
  $\GF{q}$.
\end{Proposition}

\begin{Proof}
  The proof is given in Sec.~\ref{sec:proof:prop:udms:construction:1}. Note
  that we use Cor.~\ref{cor:prop:udms:construction:1} (see below) which
  presents a reformulation of the construction.
\end{Proof}

\begin{Corollary}
  \label{cor:prop:udms:construction:1}

  Let us associate the information polynomial $u(\iL) \defeq \sum_{k \in [K]}
  u_k \iL^k \in \GF{q}[\iL]$, where $u_k \defeq [\vu]_k$, $k \in [K]$, to the
  information vector $\vu$. The construction in the above proposition results
  in a coding scheme where the vector $\vu$ is mapped to the vectors $\vx_0,
  \ldots, \vx_{L-1}$ with entries
  \begin{align*}
    [\vx_{\ell}]_n
      &= u^{(n)}(\beta_{\ell}),
         \quad (\ell,n) \in [L] \times [N],
  \end{align*}
  where $\beta_0 \defeq 0$, $\beta_1 \defeq \infty$, $\beta_{\ell+2} \defeq
  \alpha^{\ell}$, $\ell \in [L-2]$. (Note that because $\alpha$ is a primitive
  element of $\GF{q}$, all $\beta_{\ell}$, $\ell \in [L]$, are distinct.) This
  means that over the $\ell$-th channel we are transmitting the coefficients
  of the Taylor polynomial expansion of $u(\iL)$ around $\iL = \beta_{\ell}$.
\end{Corollary}

\begin{Proof}
  Follows from the paragraph after
  Def.~\ref{def:taylor:polynomial:expansion:1}.
\end{Proof}

\begin{Example}
  \label{ex:udms:construction:1}

  For $N \defeq 3$, $K \defeq N$, $p \defeq 3$, and $\alpha \defeq 2$, we
  obtain the $L = 3 + 1 = 4$ matrices that were shown in
  Ex.~\ref{ex:udm:2}. Note that $\matr{A}_3$ is nearly the same as
  $\matr{A}_2$: it differs only in that the main diagonal is multiplied by
  $\alpha^0 = 1$, the first upper diagonal is multiplied by $\alpha^1 = 2$,
  the second upper diagonal is multiplied by $\alpha^2 = 1$, the first lower
  diagonal is multiplied by $\alpha^{-1} = 2$, and the second lower diagonal
  is multiplied by $\alpha^{-2} = 1$.
\end{Example}

We collect some remarks about the UDMs constructed in
Prop.~\ref{prop:udms:construction:1} and
Cor.~\ref{cor:prop:udms:construction:1}.
\begin{itemize}

  \item All matrices $\matr{A}_{\ell}$, $2 \leq \ell < L$, are upper
    triangular matrices with non-zero diagonal entries. This follows from the
    fact that ${k \choose n} = 1$ if $k = n$ and ${k \choose n} = 0$ if $k <
    n$.

  \item The matrix $\matr{A}_2$ is an upper triangular matrix where the
    non-zero part equals Pascal's triangle (modulo $p$), see e.g.~$\matr{A}_2$
    in Ex.~\ref{ex:udm:2}. However, whereas usually Pascal's triangle is
    depicted such that the \emph{rows} correspond to the upper entry in the
    binomial coefficient, here the \emph{columns} of the matrix correspond to
    the upper entry in the binomial coefficient.

  \item Applying Lemma~\ref{lemma:reduction:1} to $(L,N,K{=}N,q)$-UDMs as
    constructed in Prop.~\ref{prop:udms:construction:1} and yields $(L,N{-}1,$
    $K{-}1{=}N{-}1,q)$-UDMs as constructed in
    Prop.~\ref{prop:udms:construction:1}.

  \item As already mentioned in Sec.~\ref{sec:introduction:1}, the
    construction of UDMs in Prop.~\ref{prop:udms:construction:1} and
    Cor.~\ref{cor:prop:udms:construction:1} is essentially equivalent to codes
    presented by Rosenbloom and
    Tsfasman~\cite{Rosenbloom:Tsfasman:97:1}.\footnote{Note that the
    communication system mentioned in Sec.~1
    of~\cite{Rosenbloom:Tsfasman:97:1} also talks about parallel channels:
    however, that communication system would correspond to (in our notation)
    sending $L$ symbols over $N$ channels. On the other hand, the
    communication system that is mentioned in
    Nielsen~\cite[Ex.~18]{Nielsen:00:1} is more along the lines of the
    Tavildar-Viswanath channel model~\cite{Tavildar:Viswanath:05:2} mentioned
    in Sec.~\ref{sec:introduction:1}.} They were interested in the so-called
    $m$-metric which is now also known as the Rosenbloom-Tsfasman
    metric.\footnote{In the context of uniform distributions, this metric was
    then later on introduced independently by Martin and
    Stinson~\cite{Martin:Stinson:99:1} and by
    Skriganov~\cite{Skriganov:01:1}.} Later, Nielsen~\cite{Nielsen:00:1}
    discussed Sudan-type decoding algorithms for these codes. Related work on
    codes under the Rosenbloom-Tsfasman metric include: Dougherty and
    Skriganov~\cite{Dougherty:Skriganov:02:1} on MacWilliams duality,
    Dougherty and Skriganov~\cite{Dougherty:Skriganov:02:1} and Dougherty and
    Shiromoto~\cite{Dougherty:Shiromoto:04:1} on codes over rings and other
    generalized alphabets, Lee~\cite{Lee:03:1} on automorphisms that preserve
    the Rosenbloom-Tsfasman metric, Chen and
    Skriganov~\cite{Chen:Skriganov:02:1} on codes with large distances
    simultaneously in the Hamming and in the Rosenbloom-Tsfasman metric.

  \item There is also a connection between the construction of UDMs in
    Prop.~\ref{prop:udms:construction:1} and
    Cor.~\ref{cor:prop:udms:construction:1} and so-called repeated-root cyclic
    codes~\cite{Chen:69:1, Castagnoli:Massey:Schoeller:vonSeemann:91:1,
    Lint:91:1, MorelosZaragoza:91:1}, i.e.~the mathematics behind both of them
    is very similar. Repeated-root cyclic codes are cyclic codes where the
    generator polynomial has zeros with multiplicity possibly larger than one:
    Lemma~\ref{lemma:zero:multiplicity:1}, the lemma that is crucial for
    proving the UDMs property for the UDMs constructed in
    Prop.~\ref{prop:udms:construction:1} and
    Cor.~\ref{cor:prop:udms:construction:1}, was used by Castagnoli et
    al.~\cite{Castagnoli:Massey:Schoeller:vonSeemann:91:1} to formulate
    parity-check matrices for repeated-root cyclic codes.

\end{itemize}

Interestingly, for $K = N$ the construction in
Prop.~\ref{prop:udms:construction:1} and
Cor.~\ref{cor:prop:udms:construction:1} is unique in a sense made more precise
in the following lemma.

\begin{Proposition}
  \label{prop:udms:construction:uniqueness:1}

  The construction in Prop.~\ref{prop:udms:construction:1} is unique in the
  following sense. Let $N$ be some positive integer, let $q$ be some prime
  power, let $\alpha$ be a primitive element in $\GF{q}$, and let $L$ be a
  positive integer with $L \leq q+1$. Moreover, let $\matr{A}_0, \ldots,
  \matr{A}_{L-1}$ be $(L,N,K{=}N,q)$-UDMs as given by
  Prop.~\ref{prop:udms:construction:1} and
  Cor.~\ref{cor:prop:udms:construction:1}. (Note that the $L \times N$ matrix
  consisting of the zeroth rows of $\matr{A}_0, \ldots, \matr{A}_{L-1}$ is the
  generator matrix of a Reed-Solomon code of length $L$ and dimension
  $\min(N,L)$.)

  Consider the $N \times N$ matrices $\matr{A}'_0, \ldots, \matr{A}'_{L-1}$
  over $\GF{q}$ such that $\matr{A}'_0 \defeq \matr{A}_0$, $\matr{A}'_1 \defeq
  \matr{A}_1$, and where the zeroth row of $\matr{A}'_{\ell}$ matches the
  zeroth row of $\matr{A}_{\ell}$ for all $2 \leq \ell \leq L-1$. Modulo the
  modifications described in Lemma~\ref{lemma:replace:and:add:lines:1}, the
  only way to fill the remaining entries of the matrices $\matr{A}'_0, \ldots,
  \matr{A}'_{L-1}$ such that they are $(L,N,K{=}N,q)$-UDMs is to choose
  $\matr{A}'_{\ell} = \matr{A}_{\ell}$ for all $2 \leq \ell \leq L-1$.
\end{Proposition}

\begin{Proof}
  See Sec.~\ref{sec:proof:prop:udms:construction:uniqueness:1}.
\end{Proof}

\mbox{}

The above proposition says something about the uniqueness of UDMs if one bases
the construction of UDMs on Reed-Solomon codes. The question is then how
unique are Reed-Solomon codes in the class of MDS codes. In that respect,
MacWilliams and Sloane~\cite[p.~330]{MacWilliams:Sloane:77} note that if $q$
is odd then in many (conjecturally all) cases there is an unique $[q{+}1, k,
q{-}k+{2}]$ $q$-ary MDS code. But if $q$ is even this is known to be false.

\begin{Corollary}
  \label{cor:udms:construction:2}

  Consider the setup of Prop.~\ref{prop:udms:construction:1} with $N = K =
  p^m$, where $p$ is the characteristic of $\GF{q}$ and where $m$ is some
  positive integer.\footnote{The statement in this corollary could be extended
  to more general setups but we will not do so.}  Let
  \begin{alignat*}{2}
    n
      &= n_{m-1} p^{m-1} + \cdots + n_1 p + n_0,
    \quad
    &
    0 \leq n_{h} < p, \
    &
    h \in [m]
    \quad \text{and} \\
    k
      &= k_{m-1} p^{m-1} + \cdots + k_1 p + k_0,
    \quad
    &
    0 \leq k_{h} < p, \
    &
    h \in [m]
  \end{alignat*}
  be the radix-$p$ representations of $n \in [N]$ and $k \in [N]$,
  respectively. Then the entries of $\matr{A}_{\ell+2}$, $\ell \in [L-2]$, can
  be written as
  \begin{align*}
    [\matr{A}_{\ell+2}]_{n,k}
      &= \prod_{h \in [m]}
           {k_{h} \choose n_{h}}
           \alpha^{\ell (k_{h} - n_{h}) p^{h}}.
  \end{align*}
  This shows that the matrices $\matr{A}_{\ell}$, $\ell \in [L]$, can be
  written as tensor products of some $p \times p$ matrices. In the special
  case $q = p$ (i.e.~$q$ is a prime) we can say more. Namely, letting
  $\matr{A}'_0, \ldots, \matr{A}'_{L-1}$ be the $(p{+}1,p,p,p)$-UDMs as
  constructed in Prop.~\ref{prop:udms:construction:1} we see that
  $\matr{A}_{\ell} = (\matr{A}'_{\ell})^{\otimes m}$ for all $\ell \in [L]$.
\end{Corollary}

\begin{Proof}
  See Sec.~\ref{sec:proof:cor:udms:construction:2}.
\end{Proof}

\mbox{}

Consider the same setup as in Cor.~\ref{cor:udms:construction:2}. Because $0
\leq n_h < p$, we observe that ${k_{h} \choose n_{h}}$ is a polynomial
function of degree $n_{h}$ in $k_h$. Using
Lemma~\ref{lemma:replace:and:add:lines:1}, the matrices can therefore be
modified so that the entries are
\begin{align*}
  [\matr{A}_{\ell+2}]_{n,k}
    &= \prod_{h \in [m]}
         k_h^{\,n_{h}} \alpha^{\ell (k_{h} - n_{h}) p^{h}},
         \quad
         (\ell,n,k) \in [L-2] \times [N] \times [N].
\end{align*}
Letting $q = p \defeq 2$, $L \defeq q + 1 = 3$, $N = K = 2^m$, and $\alpha
\defeq 1$ we have $[\matr{A}_2]_{n,k} = \prod_{h \in [m]} k_h^{\,\,n_{h}}$,
which recovers the $(L{=}3,N{=}2^m,K{=}N,q{=}2)$-UDMs
in~\cite[Sec.~4.5.3]{Tavildar:Viswanath:05:2} since the latter matrix is a
Hadamard matrix. In general (i.e.~not just in the case $q = 2$), the fact that
the entries of $[\matr{A}_2]_{n,k}$ can be written as $[\matr{A}_2]_{n,k} =
\prod_{h \in [m]} k_{h}^{n_{h}}$, reminds very strongly of Reed-Muller
codes~\cite{MacWilliams:Sloane:77, Blahut:02:1}. However, whereas in the
former case the rows of $\matr{A}_2$ are the evaluation of the multinomial
function $(t_0, \ldots, t_{m-1}) \mapsto \prod_{h \in [m]} k_{h}^{n_{h}}$, in
the latter case the columns of an $N \times K$ generator matrix of an $[N,K]$
$q$-ary Reed-Muller code can be seen as the evaluation of multinomials at
various places.

Recall the $(L{=}4,N{=}3,K{=}N,q{=}3)$-UDMs $\matr{A}_0, \ldots, \matr{A}_3$
from Ex.~\ref{ex:udm:2}. The authors of~\cite{Tavildar:Viswanath:05:2,
Doshi:05:1} conjecture that the tensor powers $\matr{A}_0^{\otimes m}, \ldots,
\matr{A}_3^{\otimes m}$ are $(4,3^m,3^m,3)$-UDMs for any positive integer
$m$. This is indeed the case and can be shown as follows. From
Ex.~\ref{ex:udms:construction:1} we know that $\matr{A}_0, \ldots,
\matr{A}_3$ can be obtained by the construction in
Prop.~\ref{prop:udms:construction:1}. Because $q = 3$ is a prime,
Cor.~\ref{cor:udms:construction:2} yields the desired conclusion that the
tensor powers $\matr{A}_0^{\otimes m}, \ldots, \matr{A}_3^{\otimes m}$ are
$(4,3^m,3^m,3)$-UDMs for any positive integer $m$.

Before concluding this section on constructions of UDMs, let us mention that
the setup in Def.~\ref{def:udms:1} can be generalized as follows: instead of
requiring that decoding is uniquely possible for any $(\us_0, \ldots,
\us_{L-1}) \in \usSet^{\geq K}_{L,N}$ one may ask that decoding is uniquely
possible for any $(\us_0, \ldots, \us_{L-1}) \in \usSet^{\geq K'}_{L,N}$ where
$K' \geq K$. Of course, UDMs designed for $K' = K$ can be used for any $K'
\geq K$, however, for suitably chosen UDMs the required field size might be
smaller, i.e.~the upper bounds on $L$ in terms of $q$ as
in~Lemmas~\ref{lemma:maximal:L:1} and~\ref{lemma:maximal:L:2} might not be a
necessary condition anymore. Indeed, in the same way as Goppa codes /
algebraic-geometry codes (see e.g.~\cite{Stichtenoth:93:1}) are
generalizations of Reed-Solomon codes, one can construct UDMs that are
generalizations of the UDMs in Prop.~\ref{prop:udms:construction:1} and
Cor.~\ref{cor:prop:udms:construction:1}; we refer to
to~\cite{Rosenbloom:Tsfasman:97:1, Nielsen:00:1} for such generalizations. The
main idea is to evaluate the information polynomial at the rational places of
a projective, geometrically irreducible, non-singular algebraic curve of genus
$g \leq K' - K$.  The proof for this setup is very similar to the proof of
Cor.~\ref{cor:prop:udms:construction:1}, however instead of the fundamental
theorem of algebra one needs the Riemann-Roch
theorem~\cite{Stichtenoth:93:1}. The Hasse-Weil-Serre
bound~\cite{Stichtenoth:93:1} (or better bounds than the Hasse-Weil-Serre
bound, cf.~e.g.~\cite{vanderGeer:vanderVlugt:05:1}) give an idea what $L$'s
can be achieved with this algebraic-geometry-based construction. An
interesting avenue for investigation is also to find upper bounds on $L$ as a
function of $N$, $K$, and $K'$ that generalize the results in
Lemmas~\ref{lemma:maximal:L:1} and~\ref{lemma:maximal:L:2} and to see if the
algebraic-geometry-based construction can achieve these bounds or if one needs
different constructions.

\section{Decoding}
\label{sec:decoding:1}

In Sec.~\ref{sec:introduction:1} we mentioned that finding $\vect{\hat u}$ was
equivalent to solving $\matr{A} \cdot \vect{\hat u} = \vect{y}$. We remind the
reader that $\matr{A}$ is the $(\sum_{\ell \in [L]}
\us_{\ell}) \times K$-matrix $\matr{A}$ that stacks the $\us_0$ first rows of
$\matr{A}_0$, $\ldots$, the $\us_{L-1}$ first rows of $\matr{A}_{L-1}$ and
that $\vy$ is the length-$(\sum_{\ell \in [L]} \us_{\ell})$ vector $\vect{y}$
that concatenates the $\us_0$ first entries of $\matr{y}_0$, $\ldots$, the
$\us_{L-1}$ first entries of $\matr{A}_{L-1}$. It is clear that if this linear
equation system contains more than $K$ equalities then we can neglect all but
the first $K$ equalities. To solve the resulting system of linear equations we
can use Gaussian elimination which results in a complexity of $O(K^3)$.

However, for specific constructions of UDMs we can do better, in particular
for UDMs as constructed in Prop.~\ref{prop:udms:construction:1} and
Cor.~\ref{cor:prop:udms:construction:1}. Assume that $(\us_0, \ldots,
\us_{L-1}) \in \usSet^{= K}_{L,N}$. (If $(\us_0, \ldots, \us_{L-1}) \in
\usSet^{\geq K}_{L,N} \setminus \usSet^{= K}_{L,N} = \setminus 
\usSet^{> K}_{L,N}$ then we can reduce some $\us_{\ell}$'s such that 
$(\us_0, \ldots, \us_{L-1}) \in \usSet^{= N}_L$.) For these UDMs, decoding
means to find the polynomial $u(\iL) = \sum_{k \in [K]} u_k \iL^k \in
\GF{q}[\iL]$ such that $u^{(n)}(\beta_{\ell}) = [\vect{y}_{\ell}]_n$, $\ell
\in [L]$, $n \in [\us_{\ell}]$, i.e.~such that $\deg(u(\iL)) < K - 1 - \us_1$
and such that $u^{(n)}(\beta_{\ell}) = [\vect{y}_{\ell}]_n$, $\ell \in [L]
\setminus \{ 1 \}$, $n \in [\us_{\ell}]$. The following algorithm, which is
based on Newton interpolation~\cite{Sokolnikoff:Redheffer:66:1}, finds the
coefficients of the polynomial $u(\iL)$ in time $O(K^2)$.

\begin{Algorithm}
  \label{alg:decoding:1}

  Let $N$ be some positive integer, let $q$ be some prime power, let $\alpha$
  be a primitive element in $\GF{q}$, and let $L$ be a positive integer with
  $L \leq q+1$. Moreover, let $\matr{A}_0, \ldots, \matr{A}_{L-1}$ be
  $(L,N,K{=}N,q)$-UDMs as given by Prop.~\ref{prop:udms:construction:1} and
  Cor.~\ref{cor:prop:udms:construction:1}. For $(\us_0, \ldots, \us_{L-1}) \in
  \usSet^{= K}_{L,N}$, the following steps find the coefficients of $u(\iL) =
  \sum_{k \in [K]} u_k \iL^k$ based on the knowledge of $[\vy_{\ell}]_n$,
  $\ell \in [L]$, $n \in [\us_{\ell}]$. \begin{itemize}

    \item Set $h := 0$ and $g(\iL) := 1$.

    \item For $\ell$ from $0$ to $L-1$ (without $1$) do
      \begin{itemize}

        \item For $n$ from 0 to $\us_{\ell} - 1$ do
          \begin{align}
            \delta
              &:= [\vect{y}_{\ell}]_n
                  -
                  h^{(n)}(\beta_{\ell}),
                    \label{eq:dec:discrepancy:1} \\
            h(\iL)
              &:= h(\iL)
                  +
                  \frac{\delta}{g^{(n)}(\beta_{\ell})}
                  \cdot
                  g(\iL),
                    \label{eq:dec:update:1} \\
            g(\iL)
              &:= g(\iL)
                  \cdot
                  (\iL - \beta_{\ell})
                    \label{eq:dec:update:2}
          \end{align}
      \end{itemize}

    \item $u_k := h_k$ for $k \in [K]$.

  \end{itemize}
\end{Algorithm}

\begin{Proof}
  See Sec.~\ref{sec:proof:alg:decoding:1}.
\end{Proof}

\section{Conclusions}
\label{sec:conclusions:1}

For $K \leq 2N$ we have presented an explicit construction of UDMs for all
parameters $L$, $N$, $K$, and $q$ for which UDMs can potentially exist. We
have pointed out connections to codes by Rosenbloom and Tsfasman, to
Reed-Solomon codes, and repeated-root cyclic codes. We have also shown in what
sense these UDMs are an unique extension of Reed-Solomon codes in the case $K
= N$. Moreover, we have presented an efficient decoding algorithm. The
construction works also for the case $K > 2N$, however it is yet not clear for
what $L$ and $q$ such UDMs can exist. Moreover, generalizing the setup as
indicated at the end of Sec.~\ref{sec:explicit:construction:udms:1}, i.e.~to
require unique decodability only for $(\us_0, \ldots,
\us_{L-1}) \in \usSet^{\geq K'}_{L,N}$ where $K' \geq K$, raises many new
questions on the existence of UDMs and on how to explicitly construct them.

\newpage

\appendix

\section{Proofs}

\label{sec:proofs:1}

\subsection{Proof of Lemma~\ref{lemma:reversed:order:2}}

\label{sec:proof:lemma:reversed:order:2}

We distinguish two cases: $K \geq 2N$ and $K \leq 2N-1$. Consider the first
case, i.e.~$K \geq 2N$. We can set $\matr{A}'_{\ell} \defeq
\matr{A}_{\ell}$ for all $\ell \in [L]$ since there is no row for which we
have to prove that it equals the row of another matrix.

So, consider now the second case, i.e.~$K \leq 2N-1$. It is sufficient to show
how $\matr{A}_0$ and $\matr{A}_1$ can be used to construct matrices
$\matr{A}'_0$ and $\matr{A}'_1$ such that $\matr{A}'_0, \matr{A}'_1,
\matr{A}_2, \ldots, \matr{A}_{L-1}$ are $(L,N,K,q)$-UDMs and such that the
$n$-th row of $\matr{A}'_1$ equals the $(K{-}1{-}n)$-th row of $\matr{A}'_0$
for all $K-N \leq n \leq N-1$. We use the following algorithm:
\begin{itemize}

  \item Assign $\matr{A}'_0 := \matr{A}_0$ and $\matr{A}'_1 := \matr{A}_1$.

  \item For $n$ from $K-N$ to $N-1$ do
    \begin{itemize}

      \item Let $\matr{B}'_0$ be the $(n+1) \times K$ matrix that contains the
        rows $0$ to $n$ from $\matr{A}'_0$. Similarly, let $\matr{B}'_1$ be
        the $(K{-}n) \times K$ matrix that contains the rows $0$ to $K-1-n$
        from $\matr{A}'_0$.

      \item Build the $(K{+}1) \times K$-matrix $\matr{B}$ by stacking
        $\matr{B}'_0$ and $\matr{B}'_1$.

      \item Because of the size of $\matr{B}$, the left null space of
        $\matr{B}$ is non-empty. (In fact, because of the UDMs conditions the
        matrix $\matr{B}$ must have rank $K$ which implies that the left null
        space is one-dimensional.) Pick a non-zero (row) vector $\vect{b}^\tr$
        in this left null space, i.e.~$\vect{b}^\tr$ fulfills $\vect{b}^\tr
        \cdot \matr{B} = \vect{0}^\tr$. Write $\vect{b}^\tr = (
        {\vect{b}'_0}^\tr \, | \, {\vect{b}'_1}^\tr)$ where $\vect{b}'_0$ is
        of length $n+1$ and $\vect{b}'_1$ is of length $K-n$.

      \item Because of the UDMs conditions it can be seen that neither
        $[\vect{b}]_n$ nor $[\vect{b}]_{K+1}$ can be zero, i.e.~neither the
        last component of $\vect{b}'_0$ nor the last component of
        $\vect{b}'_1$ is zero. Replace the $n$-th row of matrix $\matr{A}'_0$
        by the vector ${\vect{b}'_0}^\tr \matr{B}'_0$. Similarly, replace the
        $(K{-}1{-}n)$-th row of matrix $\matr{A}'_1$ by the vector
        $-{\vect{b}'_1}^\tr \matr{B}'_1$. We see that the $n$-th row of
        $\matr{A}'_0$ equals the $(K{-}1{-}n)$-th row of $\matr{A}'_1$ and
        because of Lemma~\ref{lemma:replace:and:add:lines:1} the matrices
        $\matr{A}'_0, \matr{A}'_1, \matr{A}_2, \ldots, \matr{A}_{L-1}$ are
        still $(L,N,K,q)$-UDMs.

    \end{itemize}

\end{itemize}
Applying the algorithm to $\matr{A}_2$ and $\matr{A}_3$, then to $\matr{A}_4$
and $\matr{A}_5$, $\ldots$ yields the desired result.

\subsection{Proof of Remark~\ref{remark:identity:structure:1}}

\label{sec:proof:remark:identity:structure:1}

Indeed, if $\matr{A}_0$ and $\matr{A}_1$ are not of this form then the
algorithm in the proof of Lemma~\ref{lemma:reversed:order:2} allows us to
replace these two matrices by two matrices where the $n$-th row of
$\matr{A}_0$ is the same as the $(K{-}1{-}n)$-th row of $\matr{A}_1$ for $K-N
\leq n \leq N-1$. To proceed, we distinguish two cases, $K \geq 2N$ and $K
\leq 2N-1$.

If $K \geq 2N$ then we construct a $K \times K$-matrix $\matr{B}$ as follows:
for $0 \leq n \leq N-1$ the $n$-th row of $\matr{B}$ equals the $n$-th row of
$\matr{A}_0$ and for $0 \leq n \leq N-1$ the $(K{-}1{-}n)$-th row of
$\matr{B}$ equals the $n$-th row of $\matr{A}_1$. Because of the UDMs property
it is possible to fill the unspecified rows of $\matr{B}$ such that $\matr{B}$
is an invertible matrix.

If $K \leq 2N-1$ then we construct a $K \times K$-matrix $\matr{B}$ as
follows: for $0 \leq n \leq N-1$ the $n$-th row of $\matr{B}$ equals the
$n$-th row of $\matr{A}_0$ and for $0 \leq n \leq K-N-1$ the $(K{-}1{-}n)$-th
row of $\matr{B}$ equals the $n$-th row of $\matr{A}_1$. Because of the UDMs
properties the matrix $\matr{B}$ is an invertible matrix.

Finally, for all $\ell \in [L]$ we replace the matrix $\matr{A}_{\ell}$ by the
matrix $\matr{A}_{\ell} \cdot \matr{B}^{-1}$ and we obtain the desired result.

\subsection{Proof of Lemma~\ref{lemma:reduction:1}}

\label{sec:proof:lemma:reduction:1}

It is clear that $\matr{A}'_0 = \matr{I}_{N-1,K-1}$ and $\matr{A}'_1 =
\matr{J}_{N-1,K-1}$. We know that for any $(\us_0, \ldots, \us_{L-1}) \in
\usSet^{= K}_{L,N}$ the UDMs condition is fulfilled for the matrices
$\matr{A}_0, \ldots, \matr{A}_{L-1}$. We have to show that for any $(\us'_0,
\ldots, \us'_{L-1}) \in \usSet^{= K-1}_{L,N-1}$ the UDMs condition is also
fulfilled for the matrices $\matr{A}'_0, \ldots, \matr{A}'_{L-1}$.

Take such an $L$-tuple $(\us'_0, \ldots, \us'_{L-1}) \in \usSet^{=
N-1}_{L,K-1}$. The $(K{-}1) \times (K{-}1)$-matrix $\matr{A}'$ for which we
have to check the full-rank condition looks like
\begin{align*}
  \matr{A}'
    &= \begin{pmatrix}
         \matr{I}_{\us'_0}     & \matr{0}   & \matr{0} \\
         \matr{0}              & \matr{0}   & \matr{J}_{\us'_1} \\
         \matr{B}'             & \matr{B}'' & \matr{B}'''
       \end{pmatrix},
\end{align*}
where $\matr{B}'$, $\matr{B}''$, and $\matr{B}'''$ are matrices of size
$(K{-}1{-}\us'_0{-}\us'_1) \times \us'_0$, $(K{-}1{-}\us'_0{-}\us'_1) \times
(K{-}1{-}\us'_0{-}\us'_1)$, and $(K{-}1{-}\us'_0{-}\us'_1) \times \us'_1$,
respectively, and where $[\matr{B}', \matr{B}'', \matr{B}''']$ consists of
rows from $\matr{A}'_{\ell}$, $2 \leq \ell < L$. It can easily be seen that
the $(N{-}1) \times (N{-}1)$-matrix $\matr{A}'$ has full rank if and only if
the $K \times K$-matrix
\begin{align*}
  \matr{A}
    &= \begin{pmatrix}
         \matr{I}_{\us'_0} & \matr{0}   & \matr{0}        & \vect{0} \\
         \matr{0}        & \matr{0}   & \matr{0}        & 1 \\
         \matr{0}        & \matr{0}   & \matr{J}_{\us'_1} & \vect{0} \\
         \matr{B}'       & \matr{B}'' & \matr{B}'''     & \vect{b}
       \end{pmatrix}
     = \begin{pmatrix}
         \matr{I}_{\us'_0} & \matr{0}   & \matr{0} \\
         \matr{0}        & \matr{0}   & \matr{J}_{\us'_1+1} \\
         \matr{B}'       & \matr{B}'' & \matr{B}''''
       \end{pmatrix}
\end{align*}
has full rank, where $\vect{b}$ is an arbitrary
length-$(K{-}1{-}\us'_0{-}\us'_1)$ vector and where $\matr{B}'''' \defeq
[\matr{B}''' \ | \ \vect{b}]$.

Let $\us_{\ell} \defeq \us'_{\ell}$ for $\ell \in [L] \setminus \{ 1 \}$ and
let $\us_1 \defeq \us'_1 + 1$. Clearly, $(\us_0, \ldots, \us_{L-1}) \in
\usSet^{= N}_{L,K}$. Choosing $\vect{b}$ such that the first $\us_2$ entries 
of $\vect{b}$ equal the top $\us_2$ entries of the $(K{-}1)$-th column of
$\matr{A}_2$, $\ldots$, the last $\us_{L-1}$ entries of $\vect{b}$ equal the
top $\us_{L-1}$ entries of the $(K{-}1)$-th column of $\matr{A}_{L-1}$, we see
that $\matr{A}$ represents the matrix that we have to look at when checking
the UDMs property for $(\us_0, \ldots, \us_{L-1})$ for $\matr{A}_0, \ldots,
\matr{A}_{L-1}$. However, by assumption we know that $\matr{A}$ has full rank
and so the matrix $\matr{A}'$ has also full rank.

\emph{Comment:} from this proof we see that we did not really need the
condition that $\matr{A}_0 = \matr{I}_{N,K}$ and that $\matr{A}_1 =
\matr{J}_{N,K}$, the only property of $\matr{A}_0$ and $\matr{A}_1$ that we
used was that the zeroth row of $\matr{A}_1$ is $(0, \ldots, 0, 1)$ and/or the
last column of $\matr{A}_1$ is $(1, 0, \ldots, 0)^\tr$.

\subsection{Proof of Lemma~\ref{lemma:maximal:L:1}}

\label{sec:proof:lemma:maximal:L:1}

Let $\matr{A}_0, \ldots, \matr{A}_{L-1}$ be $(L,N,K{\leq}2N,q)$-UDMs, let $k
\defeq k_0 \defeq \lfloor (K-1) / 2 \rfloor$, and let $k_1 \defeq \lfloor
(K-2) / 2 \rfloor$. (Note that $k_0 + k_2 = K - 2$.)
Rem.~\ref{remark:identity:structure:1} allows us to assume without loss of
generality that $\matr{A}_0 = \matr{I}_{N,K}$ and that $\matr{A}_1 =
\matr{J}_{N,K}$.

First, for all $2 \leq \ell \leq L-1$ we want to show that the entries
$[\matr{A}_{\ell}]_{0,k}$ and $[\matr{A}_{\ell}]_{0,k+1}$ must both be
non-zero. Indeed, the UDMs condition for $\us_0 = k_0$, $\us_1 = k_1 + 1$, and
$\us_{\ell} = 1$ (all other $\us_{\ell'}$ are zero) shows that
$[\matr{A}_{\ell}]_{0,k} \neq 0$, and the UDMs condition for $\us_0 = k_0 +
1$, $\us_1 = k_1$, and $\us_{\ell} = 1$ (all other $\us_{\ell'}$ are zero)
shows that $[\matr{A}_{\ell}]_{0,k+1} \neq 0$. Using
Lemma~\ref{lemma:replace:and:add:lines:1} we can therefore without loss of
generality assume that $[\matr{A}_{\ell}]_{0,k} = 1$ for all $2 \leq
\ell < L$.

Secondly, the UDMs condition for $\us_0 = k_0$, $\us_1 = k_1$, $\us_{\ell}
= 1$, and $\us_{\ell'} = 1$ (all other $\us_{\ell''}$ are zero) implies that
the matrix
\begin{align*}
  \begin{pmatrix}
    [\matr{A}_{\ell}]_{0,k}  & [\matr{A}_{\ell}]_{0,k+1} \\
    [\matr{A}_{\ell'}]_{0,k} & [\matr{A}_{\ell'}]_{0,k+1}
  \end{pmatrix}
    &= \begin{pmatrix}
         1 & [\matr{A}_{\ell}]_{0,k+1}  \\
     1 & [\matr{A}_{\ell'}]_{0,k+1}
       \end{pmatrix}
\end{align*}
must have rank $2$ for any distinct $\ell$ and $\ell'$ fulfilling $2 \leq \ell
< L$ and $2 \leq \ell' < L$. It is not difficult to see that this implies that
$[\matr{A}_{\ell}]_{0,k+1}$ must be distinct for all $2 \leq \ell < L$. Since
$[\matr{A}_{\ell}]_{0,k+1}$ must be non-zero and since $\GF{q}$ has $q-1$
non-zero elements we see that $L-2 \leq q-1$, i.e.~$L \leq q+1$.

\subsection{Proof of Lemma~\ref{lemma:maximal:L:2}}

\label{sec:proof:lemma:maximal:L:2}

Let $\matr{A}_0, \ldots, \matr{A}_{L-1}$ be $(L,N,K{=}2N{+}1,q)$-UDMs.
Rem.~\ref{remark:identity:structure:1} allows us to assume without loss of
generality that $\matr{A}_0 = \matr{I}_{N,2N+1}$ and that $\matr{A}_1 =
\matr{J}_{N,2N+1}$.  For all $2 \leq \ell < L$ the UDMs condition for $\us_0 =
N$, $\us_1 = N$, and $\us_{\ell} = 1$ (all other $\us_{\ell'}$ are zero) shows
that $[\matr{A}_{\ell}]_{0,N} \neq 0$. Using
Lemma~\ref{lemma:replace:and:add:lines:1} we can therefore without loss of
generality assume that $[\matr{A}_{\ell}]_{0,N} = 1$ for all $2 \leq
\ell < L$.

Secondly, the UDMs condition for $\us_0 = N$, $\us_1 = N-1$, $\us_{\ell}
= 1$, and $\us_{\ell'} = 1$ (all other $\us_{\ell''}$ are zero) implies that
the matrix
\begin{align*}
  \begin{pmatrix}
    [\matr{A}_{\ell}]_{0,N}  & [\matr{A}_{\ell}]_{0,N+1} \\
    [\matr{A}_{\ell'}]_{0,N} & [\matr{A}_{\ell'}]_{0,N+1}
  \end{pmatrix}
    &= \begin{pmatrix}
         1 & [\matr{A}_{\ell}]_{0,N+1}  \\
     1 & [\matr{A}_{\ell'}]_{0,N+1}
       \end{pmatrix}
\end{align*}
must have rank $2$ for any distinct $\ell$ and $\ell'$ fulfilling $2 \leq \ell
< L$ and $2 \leq \ell' < L$. It is not difficult to see that this implies that
$[\matr{A}_{\ell}]_{0,N+1}$ must be distinct for all $2 \leq \ell < L$. Since
$\GF{q}$ has $q$ elements we see that $L-2 \leq q$, i.e.~$L \leq q+2$.

\subsection{Proof of Prop.~\ref{prop:udms:construction:1}}

\label{sec:proof:prop:udms:construction:1}

We have to check the UDMs condition for all $(\us_0, \ldots, \us_{L-1}) \in
\usSet^{= K}_{L,N}$. Fix such a tuple $(\us_0, \ldots, \us_{L-1}) \in \usSet^{=
K}_{L,N}$ and let $\psi$ be the mapping of the vector $\vect{u}$ to the
non-erased entries of the vectors $\vect{y}_{\ell}$, $\ell \in [L]$; it is
clear that $\psi$ is a linear mapping. Reconstructing $\vect{u}$ is therefore
nothing else than applying the mapping $\psi^{-1}$ to the non-erased positions
of $\vect{y}_{\ell}$, $\ell \in [L]$. However, this gives an unique vector
$\vect{u}$ only if $\psi$ is an injective function. Because $\psi$ is linear,
showing injectivity of $\psi$ is equivalent to showing that the kernel of
$\psi$ contains only the vector $\vect{u} = \vect{0}$, or equivalently, only
the polynomial $u(\iL) = 0$.

So, let us show that the only possible pre-image of
\begin{align*}
  [\vect{y}_{\ell}]_n
    &= 0
       \quad
       (\ell \in [L], \ n \in [\us_{\ell}])
\end{align*}
or, equivalently, of
\begin{align*}
  [\vect{x}_{\ell}]_n
    &= 0
       \quad
       (\ell \in [L], \ n \in [\us_{\ell}])
\end{align*}
is $u(\iL) = 0$. Using the definition of $[\vect{x}_{\ell}]_n$ this is
equivalent to showing that
\begin{alignat}{2}
  u^{(n)}(\beta_{\ell})
    &= 0
       \quad
       &&(\ell \in [L] \setminus \{ 1 \}, \ n \in [\us_{\ell}])
         \label{eq:zero:hasse:derivatives:1} \\
  u_{K-1-n}
     = u^{(n)}(\beta_{\ell})
    &= 0
       \quad
       && (\ell = 1, \ n \in [\us_{\ell}])
         \label{eq:u:zero:positions:1}
\end{alignat}
implies that $u(\iL) = 0$. In a first step,
Eq.~\eqref{eq:zero:hasse:derivatives:1} and
Lemma~\ref{lemma:zero:multiplicity:1} tell us that $\beta_{\ell}$, $\ell \in
[L] \setminus \{ 1 \}$, must be a root of $u(\iL)$ of multiplicity at least
$\us_{\ell}$. Using the fundamental theorem of algebra we get
\begin{align}
  \deg(u(\iL))
    &\geq
       \sum_{\ell \in [L] \setminus \{ 1 \}}
         \us_\ell
     = K - \us_1
  \quad\quad
  \text{ or }
  \quad\quad
  u(\iL) = 0.
         \label{eq:lower:bound:number:of:roots:1}
\end{align}
In a second step, Eq.~\eqref{eq:u:zero:positions:1} tells us that we must have
$\deg(u(\iL)) \leq K - 1 - \us_1$. Combining this
with~\eqref{eq:lower:bound:number:of:roots:1}, we obtain the desired result
that $u(\iL) = 0$.

\emph{Remark:} A popular way of deriving the minimum distance of Reed-Solomon
is by using the fundamental theorem of algebra. Note however that in contrast
to the proof above, in the case of Reed-Solomon codes we do not exploit the
full potential of the fundamental theorem of algebra because there all roots
have multiplicity exactly one.

\subsection{Proof of Lemma~\ref{prop:udms:construction:uniqueness:1}}

\label{sec:proof:prop:udms:construction:uniqueness:1}

We only show that, modulo the modifications described in
Lemma~\ref{lemma:replace:and:add:lines:1}, there is an unique way of filling
the rows of $\matr{A}'_2$; the proof for the matrices $\matr{A}'_{\ell}$, $3
\leq \ell \leq L-1$ is analogous. Note that for this proof we will not work
directly with the matrix $\matr{A}'_2$ but with the mapping from $\vect{u}$ to
$\vect{x}_2$, cf.~Cor.~\ref{cor:prop:udms:construction:1}.

We know that $[\vx_2]_n = \sum_{k \in [N]} [\matr{A}'_{2}]_{n,k} u_k$ for all
$n \in [N]$.  Because of the (linear) bijection between the coefficients
$(u_k)_{k \in [N]} \in \GF{q}^N$ of $u(\iL)$ and $(u^{(n)}(\beta_2))_{n \in
[N]} \in \GF{q}^N$ (a bijection that we pointed out after
Def.~\ref{def:taylor:polynomial:expansion:1}) we can without loss of
generality assume that $\vx_2$ is obtained as follows: $[\vx_2]_n =
\sum_{j \in [N]} d_{n,j} u^{(j)}(\beta_2)$, $n \in [N]$, for some $d_{n,j}$,
$(n,j) \in [N] \times [N]$. Because we know the entries of the zeroth row of
$\matr{A}'_2$, it is clear that $d_{0,0} = 1$ and that $d_{0,j} = 0$ for $j
\in [N] \setminus \{ 0 \}$. In the remainder of the proof we will show that,
modulo the modifications described in
Lemma~\ref{lemma:replace:and:add:lines:1}, $d_{n,j} = 1$ if $n = j$ and
$d_{n,j} = 0$ otherwise, where $(n,j) \in [N] \times [N]$. This will then
imply that $\matr{A}'_2 = \matr{A}_2$.

The proof is by induction. So, for some $n \in [N]$ assume that we have shown
that $d_{n',j} = 1$ if $n' = j$ and $d_{n',j} = 0$ otherwise, where $(n',j)
\in [n] \times [N]$. This assumption is clearly fulfilled for $n = 1$, so we
only have to show that this assumption remains correct when going from line
$n$ to line $n+1$.

Let us first show that $d_{n,n} \neq 0$. Indeed, consider $(\us_0, \ldots,
\us_{L-1}) \in \usSet^{= N}_{L,N}$ with $\us_1 = N - 1 - n$ and $\us_2 = n + 1$
and all other $\us_{\ell}$ equal to zero. We know that the mapping $\psi$ from
$\vu$ to the corresponding positions of $\vect{x}_{\ell}$, $\ell
\in [L]$, is injective, i.e.~that the kernel is trivial. So, assume
that the corresponding entries of $\vect{x}_{\ell}$, $\ell \in [L]$, are zero,
i.e.~that $[\vect{x}_{\ell}]_n$, $\ell \in [L]$, $n \in [\us_\ell]$, are
zero. The special choice of $(\us_0, \ldots, \us_{L-1})$ implies that
\begin{align}
  u(\iL) &= c \cdot (\iL - \beta_2)^n, \label{eq:uniqueness:case:1:1} \\
    \sum_{j \in [N]} d_{n,j} u^{(j)}(\beta_2) &= 0,
    \label{eq:uniqueness:case:1:2}
\end{align}
for some $c \in \GF{q}$. Showing that the kernel of $\psi$ is trivial is
equal to showing that $c =
0$. Using~\eqref{eq:hasse:derivative:property:1:2} it follows
from~\eqref{eq:uniqueness:case:1:1} that $u^{(n)}(\beta_2) = c$, and that
$u^{(j)}(\beta_2) = 0$ for $j \in [N] \setminus \{ n \}$. Therefore,
\eqref{eq:uniqueness:case:1:2} reduces to $d_{n,n} \cdot c = 0$. So, for $c$
to be zero we must have $d_{n,n} \neq 0$.

Because $d_{n',n'} = 1$ for $n' \in [n]$, and using again
Lemma~\ref{lemma:replace:and:add:lines:1}, we have the freedom to set $d_{n,n}
\defeq 1$ and $d_{n,j} \defeq 0$ for $j \in [n]$. So, it remains only to show
that $d_{n,j} \defeq 0$ for $n < j < N$. We will show this by an (inner)
induction loop. Assume that for some $j$ with $n < j < N$ that we have shown
that $d_{n,j'} = 0$ for $n < j' < j$. This assumption is clearly fulfilled for
$j = n+1$ and so we only have to show that this assumption remains correct
when going from column $j$ to column $j+1$.

Let $\delta \in \{ 0, 1 \}$. Consider $(\us_0, \ldots, \us_{L-1}) \in
\usSet^{= N}_{L,N}$ with $\us_0 = j - n - \delta$, $\us_1 = N - j - 1$,
$\us_2 = n + 1$, $\us_{m} = \delta$ for some $2 < m < L$, and all other
$\us_{\ell}$ equal to zero. We know that the mapping $\psi$ from $\vu$ to the
corresponding positions of $\vect{x}_{\ell}$, $\ell
\in [L]$, is injective, i.e.~that the kernel is
trivial. So, assume that the corresponding entries of $\vect{x}_{\ell}$, $\ell
\in [L]$, are zero, i.e.~that $[\vect{x}_{\ell}]_n$, $\ell \in [L]$, 
$n \in [\us_\ell]$, are zero.  The special choice of $(\us_0, \ldots,
\us_{L-1})$ implies that
\begin{align}
  u(\iL)
    &= c \cdot (\iL-\beta_0)^{j-n-\delta}
         \cdot (\iL - \beta_2)^n
         \cdot (\iL - \beta_m)^\delta,
    \label{eq:uniqueness:case:2:1} \\
  \sum_{j' \in [N]}
    d_{n,j'} u^{(j')}(\beta_2)
    &= 0
    \label{eq:uniqueness:case:2:2}
\end{align}
for some $c \in \GF{q}$. Showing that the kernel is trivial is equal to
showing that $c = 0$. Using~\eqref{eq:hasse:derivative:property:1:1} it
follows from~\eqref{eq:uniqueness:case:2:1} that
$u^{(n)}(\beta_2) = c \cdot (\beta_2-\beta_0)^{j-n-\delta} \cdot (\beta_2 -
\beta_m)^{\delta}$, that $u^{(j)}(\beta_2) = c$, and that $u^{(j')}(\beta_2) 
= 0$ for $j < j' < N$ and so~\eqref{eq:uniqueness:case:2:2} reduces to $c
\cdot \bigl[ d_{n,n} \cdot (\beta_2 - \beta_0)^{j-n-\delta} \cdot (\beta_2 -
\beta_m)^{\delta} + d_{n,j} \bigl] = 0$. (Note that here we used the induction
assumption that $d_{n,j'} = 0$ for $n < j' < j$.) So, for $c$ to be zero we
must have
\begin{align}
  d_{n,n} \cdot (\beta_2 - \beta_0)^{j-n-\delta} \cdot (\beta_2 -
          \beta_m)^{\delta} + d_{n,j} &\neq 0.  \label{eq:uniqueness:case:2:3}
\end{align}
If $\delta = 0$ then~\eqref{eq:uniqueness:case:2:3} reduces to $d_{n,j} \neq
- d_{n,n} \cdot (\beta_2 - \beta_0)^{j-n}$ and if $\delta = 1$
then~\eqref{eq:uniqueness:case:2:3} reduces to $d_{n,j} \neq - d_{n,n} \cdot
(\beta_2 - \beta_0)^{j-n-1} \cdot (\beta_2 - \beta_m)$ for any $2 < m <
L$. This can be compactly written as $d_{n,j} \neq - d_{n,n} \cdot (\beta_2 -
\beta_0)^{j-n-1} \cdot (\beta_2 - \beta_{m'})$ for $m' \in [L] \setminus \{
1, 2 \}$. Because $d_{n,n} \cdot (\beta_2 - \beta_0)^{j-n-1}$ is non-zero and
because $\beta_2 - \beta_{m'}$ ranges over all non-zero elements of $\GF{q}$,
we see that only possibility is $d_{n,j} = 0$.

This concludes the inner induction step and therefore also the outer
induction step. During this process we have not checked all the necessary UDMs
conditions, however since we know that they all hold for $\matr{A}_0, \ldots,
\matr{A}_{L-1}$, they obviously all hold for $\matr{A}'_0, \ldots,
\matr{A}'_{L-1}$ too.

\subsection{Proof of Lemma~\ref{cor:udms:construction:2}}

\label{sec:proof:cor:udms:construction:2}

Note that ${k \choose n}$ is an integer and therefore (by the natural mapping)
an element of the prime subfield $\GF{p}$ of $\GF{q}$. Using the Lucas
correspondence theorem~\cite{Lucas:05:1} which states that ${k \choose n} =
\prod_{h \in [m]} {k_h \choose n_h}$ in $\GF{p}$ (and therefore also in
$\GF{q}$), we obtain the reformulation. The last statement in the corollary
follows from the fact that $\alpha^p = \alpha$ if $q = p$ and so $\alpha^{p^h}
= \alpha$ for any non-negative integer $h$. (Note that for $\matr{A}_{0} =
\matr{I}_{p^m}$ and $\matr{A}_{1} = \matr{J}_{p^m}$ it is trivial to verify
that they can be written as tensor product and tensor powers of $p \times p$
matrices.)

\subsection{Proof of Algorithm~\ref{alg:decoding:1}}

\label{sec:proof:alg:decoding:1}

\begin{Proof}
  Before the $m$-th execution
  of~\eqref{eq:dec:discrepancy:1}-\eqref{eq:dec:update:2} the degree of
  $g(\iL)$ is $m$. Therefore, because
  ~\eqref{eq:dec:discrepancy:1}-\eqref{eq:dec:update:2} is executed
  $\sum_{\ell \in [L] \setminus \{ 1 \}} \us_{\ell} = K - \us_1$ times, the
  degree of $h(\iL)$ is smaller than $K - \us_1$.

  The rest of the proof is by induction. We would like to show that for any
  loop variables $\ell$ and $n$ during the execution of the algorithm the
  polynomials $h(\iL)$ and $g(\iL)$ fulfill the following conditions before
  the execution of~\eqref{eq:dec:discrepancy:1}: $h^{(n')}(\beta_{\ell'}) =
  [\vect{y}_{\ell'}]_{n'}$ and $g^{(n')}(\beta_{\ell'}) = 0$ for $\ell' \in
  [\ell] \setminus \{ 1 \}$, $0 \leq n' < \us_{\ell'}$, and
  $h^{(n')}(\beta_{\ell}) = [\vect{y}_{\ell}]_{n'}$ and
  $g^{(n')}(\beta_{\ell}) = 0$ for $0 \leq n' < n$.

  It can easily be seen that this assumption holds for $\ell = 0$ and $n = 0$.
  So, consider arbitrary loop variables $\ell$ and $n$ during the execution of
  the algorithm and assume that the above-mentioned conditions
  hold. After~\eqref{eq:dec:update:1} it is clear that
  $h^{(n')}(\beta_{\ell'}) = [\vect{y}_{\ell'}]_{n'}$ for $\ell' \in [\ell]
  \setminus \{ 1 \}$, $0 \leq n' < \us_{\ell'}$, and $h^{(n')}(\beta_{\ell}) =
  [\vect{y}_{\ell}]_{n'}$ for $0 \leq n' < n$. Moreover, thanks to the
  additional term with the discrepancy factor $\delta$ we obtain
  $h^{(n)}(\beta_{\ell}) = [\vect{y}_{\ell}]_{n}$. Finally,
  after~\eqref{eq:dec:update:2} it is clear that $g^{(n')}(\beta_{\ell'}) = 0$
  for $\ell' \in [\ell] \setminus \{ 1 \}$, $0 \leq n' < \us_{\ell'}$ and
  $g^{(n')}(\beta_{\ell}) = 0$ for $0 \leq n' \leq n$. This shows that the
  above-mentioned conditions also hold at the beginning of the next execution
  of the inner loop.
\end{Proof}

\section{Hasse Derivatives}

\label{sec:hasse:derivatives:1}

Hasse derivatives were introduced in~\cite{Hasse:36:1}. Throughout this
appendix, let $q$ be some prime power. For any non-negative integer $i$, the
$i$-th Hasse derivative of a polynomial $a(\iX) \defeq \sum_{k=0}^{d} a_k
\iX^k \in \GF{q}[\iX]$ is defined to be\footnote{The $i$-th formal
derivative equals $i!$ times the Hasse derivative: so, for fields with
characteristic zero there is not a big difference between these two
derivatives since $i!$ is always non-zero, however for finite fields there can
be quite a gap between these two derivatives since $i!$ can be zero or
non-zero.}
\begin{align*}
  a^{(i)}(\iX)
    &\defeq
       \HD{\iX}{i}
         \left(
           \sum_{k=0}^{d}
             a_k \iX^k
         \right)
     \defeq
       \sum_{k=0}^{d}
         {k \choose i}  a_k \iX^{k-i}.
\end{align*}
Note that when $i > k$ then ${k \choose i} \iX^{k-i} = 0$, i.e.~the zero
polynomial. Be careful that $\HD{\iX}{i_1} \HD{\iX}{i_2} \neq
\HD{\iX}{i_1 + i_2}$ in general. However, it holds that $\HD{\iX}{i_1}
\HD{\iX}{i_2} = {i_1 + i_2 \choose i_1} \HD{\iX}{i_1 + i_2}$.

We list some well-know properties of the Hasse derivatives:
\begin{align}
  \HD{\iX}{i}
    \big(
      \gamma f(\iX) + \eta g(\iX)
    \big)
      &= \gamma \HD{\iX}{i}\big( f(\iX) \big)
         +
         \eta \HD{\iX}{i}\big( g(\iX) \big),
           \nonumber \\
  \HD{\iX}{i}
    \big(
      f(\iX) g(\iX)
    \big)
      &= \sum_{i'=0}^{i}
           \HD{\iX}{i'}\big( f(\iX) \big)
           \HD{\iX}{i-i'}\big( g(\iX) \big),
             \nonumber \\
  \HD{\iX}{i}
    \left(
      \prod_{h \in [M]}
        f_{h}(\iX)
    \right)
      &= \sum_{(i_0, \ldots, i_{M-1}) \in \usSet^{= i}_{M,i}} \
           \prod_{h \in [M]}
             \HD{\iX}{i_{h}}\big( f_{h}(\iX) \big),
               \label{eq:hasse:derivative:property:1:1} \\
  \HD{\iX}{i}
    \left(
      (X-\gamma)^k
    \right)
      &= {k \choose i} (X-\gamma)^{k-i},
           \label{eq:hasse:derivative:property:1:2}
\end{align}
where $K$ and $i$ are some non-negative integers, $M$ is some positive
integer, and where $\gamma, \eta \in \GF{q}$. The fact that a $\usSet$-set
appears in Def.~\ref{def:udms:1} and in
Eq.~\eqref{eq:hasse:derivative:property:1:1} certainly points towards the
usefulness of Hasse derivatives for constructing UDMs.

\section*{Acknowledgments}

We would like to thank Nigel Boston for general discussions on the topic, Ralf
Koetter for pointing out to us the papers~\cite{Rosenbloom:Tsfasman:97:1,
Nielsen:00:1} and for discussions on Newton interpolation, and Kamil
Zigangirov for providing us with a copy of~\cite{Rosenbloom:Tsfasman:97:1}.

\newpage

{
\bibliographystyle{amsplain}
\bibliography{/home/vontobel/references/references}
}

\end{document}